\documentstyle[12pt,epsfig]{article}        
\newlength{\dinwidth}                       
\newlength{\dinmargin}                      
\def\lsim{\mathrel{\rlap{\lower4pt\hbox{\hskip1pt$\sim$}}
    \raise1pt\hbox{$<$}}}                
\def\gsim{\mathrel{\rlap{\lower4pt\hbox{\hskip1pt$\sim$}}
    \raise1pt\hbox{$>$}}}                
\def\kf{{\bf k}}

\newcommand{\Pam}{I\!\!P}
\def\be{\begin{equation}}
\def\ee{\end{equation}}
\def\bea{\begin{eqnarray}}
\def\eea{\end{eqnarray}}
\begin{document}
\vspace*{1cm}
\begin{center}  \begin{Large} \begin{bf}
Quark-Antiquark Jets in DIS Diffractive Dissociation\\
  \end{bf}  \end{Large}
  \vspace*{5mm}
  \begin{large}
J.\ Bartels$^{\star}$, C.\ Ewerz$^{\star}$, H.\ Lotter$^{\star}$\\ 
M.\ W\"usthoff\,$^{\dagger}$\\
M.\ Diehl$^{\ddagger}$\\  
  \end{large}
\end{center}
\begin{center}
$^{\star}$II.\ Institut f\"ur Theoretische Physik, Universit\"at Hamburg,\\
     Luruper Chaussee 149,~D-22761~Hamburg,~FRG\\
$^{\dagger}$High Energy Physics Division, Argonne National Laboratory,\\
     Argonne, USA\\
$^{\ddagger}$Department of Applied Mathematics and Theoretical Physics,\\
    University of Cambridge, Cambridge CB3 9EW, England\\
\end{center}
\begin{quotation}
\noindent
{\bf Abstract:}
We report on investigations concerning the production of 
large transverse momentum
jets in DIS diffractive dissociation. 
These processes constitute a new class of events 
that allow 
for a clean test of perturbative QCD and of the hard (perturbative) 
pomeron picture. 
The measurement of the corresponding cross sections 
might possibly serve to determine the gluon density of the proton. 
\end{quotation}
\section{Introduction}
The subject of this report is the production of jets with large transverse
momenta 
in diffractive deep inelastic scattering. 
We will concentrate on events with two jets in the final state, 
i.\,e.\ events with a quark-- and an antiquark jet. 
Due to the high photon virtuality $Q^2$ and the large transverse 
momenta of the jets we can use perturbative QCD to describe this 
process. From a theoretical point of view, 
diffractive jet production should allow an even better test of pQCD 
than diffractive vector meson production since it does not involve 
the uncertainties connected with the wave function of the meson.

After defining the kinematic variables we will present in some 
detail the main results of our analysis that was performed for 
the production of light quark jets. 
A more extensive account of these results can be found 
in \cite{BLW,BELW}. 
Related and in part similar work on quark--antiquark jet 
production has been reported in \cite{Ryskin,NikZak}. 
We will close this report with a comment on open charm production. 

\section{Kinematics}
We assume the total energy $s$ to be much larger than the photon 
virtuality $Q^2=-q^2$ and much larger than the squared 
invariant mass $M^2 = (q + x_{\Pam}p)^2$ of the jet pair. 
\begin{figure}[htb]
  \begin{center}
    \leavevmode
     \input{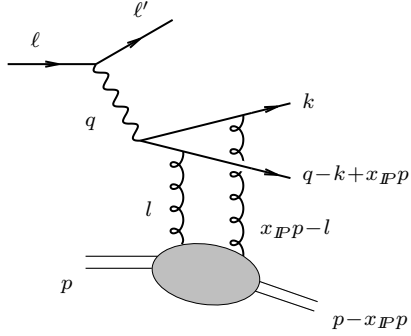}
  \end{center}
\caption{{\it
One of the four diagrams contributing to the hard scattering 
$\gamma^* + p \rightarrow q\bar{q} + p$. The outgoing (anti)quark 
momenta are held fixed.
  }}
\label{kinematics}
\end{figure}
We constrain our analysis to events with a rapidity gap and 
we require the momentum fraction $x_{\Pam}$ of the proton's 
momentum carried by the pomeron to be small, $x_{\Pam} \ll 1$. 
$x_{\Pam}$ can be expressed as 
$x_{\Pam} = (M^2 + Q^2)/(W^2 + Q^2)$ where $W$ is the invariant 
mass of the final state (including the outgoing proton). 
We keep the transverse momentum $\kf$ of the jets fixed with 
$\kf^2 \ge 1\, \mbox{GeV}^2$. 
It will be convenient to use also $\beta = x_B/x_{\Pam} = Q^2/(M^2+Q^2)$. 
The momentum transfer $t$ is taken to be zero because the cross 
section strongly peaks at this point. An appropriate $t$--dependence
taken from the elastic proton form factor is put in later by hand.

\begin{figure}[htb] 
  \begin{center}
    \leavevmode
     \input{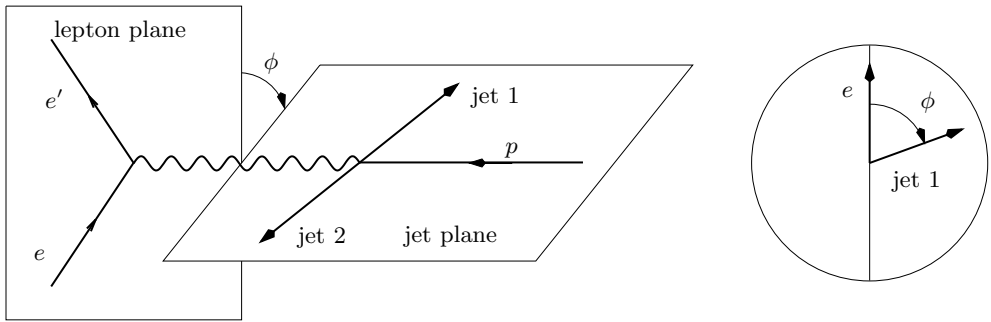}
  \end{center}
\caption{{\it
Definition of the azimuthal angle $\phi$ in the $\gamma^\ast$--$\Pam$ CMS
  }}
\label{angles}
\end{figure}

For large energy $s$ (small $x_{\Pam}$) the amplitude is dominated 
by perturbative two gluon exchange as indicated in 
fig.\ \ref{kinematics}, where the kinematic variables are 
illustrated. 
In fig.\ \ref{angles} we define the angle $\phi$ between 
the electron scattering plane and the direction of the quark jet 
pointing in the proton hemisphere (jet 1 in the figure). 
The angle $\phi$ is 
defined in the $\gamma^\ast$--$\Pam$ center of mass system 
and runs from $0$ to $2 \pi$. 

\section{Results}
In the double logarithmic approximation (DLA), 
the amplitude of the process can be expressed in terms of the 
gluon structure function. 
The cross section is therefore proportional 
to the square of the gluon structure function. 
The momentum scale of the latter can be calculated. 
We thus find as one of our main results
\be
d \sigma \sim \left[ \, x_{\Pam} G_p 
\left( x_{\Pam} ,\kf^2 \frac{Q^2+M^2}{M^2} 
\right) \right]^2 \,. 
\label{sigmagluon}
\ee
Performing our numerical estimates, however, we include some of the 
next--to--leading corrections (proportional to the momentum 
derivative of the gluon structure function) which we expect to 
be the numerically most important ones.

From (\ref{sigmagluon}) one can deduce that, in our model, 
Regge factorization \`a la Ingelman and Schlein \cite{IngSchl} 
is not valid, i.\,e.\ the cross section can {\em not} 
be written as a $x_{\Pam}$--dependent flux factor times a 
$\beta$-- and $Q^2$--dependent function.

In the following we present only 
the contribution of transversely polarized photons  
to the cross section. The corresponding plots for longitudinal 
polarization can be found in \cite{BLW,BELW}. As a rule 
of thumb, the longitudinal contribution is smaller 
by a factor of ten. Only in the region of large $\beta$ 
it becomes comparable in size. 

Figure \ref{fig:xpdep} shows the $x_{\Pam}$--dependence of the cross
section for $Q^2 = 50 \,\mbox{GeV}^2$, $\beta=2/3$ and $\kf^2 > 2
\,\mbox{GeV}^2$. We use GRV next--to--leading order parton
distribution functions \cite{GRVNL}. Our prediction is compared with
the cross section obtained in the soft pomeron model of Landshoff and
Nachtmann \cite{Diehl,LNDL} with nonperturbative two--gluon exchange.
Its flat $x_{\Pam}$--dependence, characteristic of the soft pomeron,
is quite in contrast to our prediction.

Further, we have included (indicated by {\em hybrid}) 
a prediction obtained in the framework 
of the model by M.\ W\"usthoff \cite{MarkPHD}. 
This model introduces a parametrization of the pomeron based on 
a fit to small $x_B$ data for the proton structure function $F_2$. 
In this fit, the pomeron intercept is made scale dependent in order 
to account for the transition from soft to hard regions. 
The $x_{\Pam}$--dependence is not quite as steep 
as ours but comparable in size. 
\setlength{\unitlength}{1cm}
\begin{figure}[htbp] 
\begin{center}
\begin{minipage}{6cm}
\input{fig2a.pstex_t}
\caption{{\it $x_{\Pam}$--spectrum}}
\label{fig:xpdep}
\end{minipage}
\hspace{1cm}
\begin{minipage}{6cm}
\input{fig4a.pstex_t}
\caption{{\it $\kf^2$--spectrum}}
\label{fig:ktdep}
\end{minipage}
\end{center}
\end{figure}

Figure \ref{fig:ktdep} presents the $\kf^2$--spectrum for 
different values of $Q^2$ between $15 \,\mbox{GeV}^2$ 
and $45 \,\mbox{GeV}^2$. Here we have chosen 
$x_{\Pam} = 5 \cdot 10^{-3}$ and $\beta = 2/3$. 
The quantity $\delta$ given with each $Q^2$ value describes 
the effective slope of the curves as obtained from a numerical fit 
of a power behaviour $\sim (\kf^2)^{-\delta}$. We have taken $\kf^2$ down to
0.5 GeV$^2$. For $\beta=2/3$ the effective momentum scale of the gluon
structure function in (1) equals $\kf^2 /(1-\beta)=1.5$ GeV$^2$. 

Integrating the cross section for different minimal values 
of $\kf^2$ we find that the total cross section is dominated 
by the region of small $\kf^2$. If we choose, for instance, 
$x_{\Pam} < 0.01$, $10 \,\mbox{GeV}^2 \le Q^2$ and 
$50 \,\mbox{GeV} \le W \le 220 \,\mbox{GeV}$, the total cross section 
is 
\bea
  \sigma_{\mbox{\scriptsize tot}} &=& 20\,\mbox{pb} \phantom{0}\;\;\;\;\;
   \mbox{for} \; \kf^2 \ge 5 \,\mbox{GeV}^2 \nonumber\\
  \sigma_{\mbox{\scriptsize tot}} &=& 117\,\mbox{pb} \;\;\;\;\; 
   \mbox{for} \; \kf^2 \ge 2 \,\mbox{GeV}^2  \,. \nonumber
\eea
In the hybrid model of \cite{MarkPHD} the corresponding numbers are
\bea
  \sigma_{\mbox{\scriptsize tot}} &=& 28\,\mbox{pb} \phantom{0}\;\;\;\;\;
   \mbox{for} \; \kf^2 \ge 5 \,\mbox{GeV}^2 \nonumber\\
  \sigma_{\mbox{\scriptsize tot}} &=& 108\,\mbox{pb} \;\;\;\;\; 
   \mbox{for} \; \kf^2 \ge 2 \,\mbox{GeV}^2  \,. \nonumber
\eea
In accordance with fig.\ \ref{fig:ktdep} these number show that 
the cross section is strongly suppressed with $\kf^2$. 
We are thus observing a higher twist effect here. 
For comparison, we quote the numbers which are obtained 
in the soft pomeron model \cite{Diehl}. With the same cuts the total 
cross section is
\bea
  \sigma_{\mbox{\scriptsize tot}} &=& 10.5\,\mbox{pb} \;\;\;\;\;
   \mbox{for} \; \kf^2 \ge 5 \,\mbox{GeV}^2 \nonumber\\
  \sigma_{\mbox{\scriptsize tot}} &=& 64  \,\mbox{pb} \phantom{.5}\;\;\;\;\; 
   \mbox{for} \; \kf^2 \ge 2 \,\mbox{GeV}^2  \,. \nonumber
\eea 

The $\beta$--spectrum of the cross section is shown in fig.\ 
\ref{fig:betadep} for three different values of $Q^2$. 
Here we have chosen $x_{\Pam} = 5 \cdot 10^{-3}$ 
and $\kf^2 > 2\,\mbox{GeV}^2$. 
The curves exhibit maxima which, 
for not too large $Q^2$, are located well below $\beta = 0.5$. 
For small $\beta$ we expect the production of an extra gluon to become
important. First studies in this direction have been reported 
in \cite{MarkPHD,LW}
and in \cite{Levin}, but a complete calculation has not been done yet.
\setlength{\unitlength}{1cm}
\begin{figure}[htbp] 
\begin{center}
\begin{minipage}{6cm}
\input{fig5a.pstex_t}
\caption{{\it $\beta$--spectrum}}
\label{fig:betadep}
\end{minipage}
\hspace{1cm}
\begin{minipage}{6cm}
\input{angle.pstex_t}
\caption{{\it Azimuthal angular distribution}}
\label{fig:phidep}
\end{minipage}
\end{center}
\end{figure}

The most striking observation made in \cite{BELW} concerns the 
azimuthal angular distribution, i.\,e.\ the $\phi$--distribution 
of the jets. It turns out that 
the jets prefer a plane {\em perpendicular} to the electron 
scattering plane. This behaviour comes as a surprise because 
in a boson gluon fusion process the jets appear 
dominantly {\em in} the electron scattering plane \cite{BGF}. 
The azimuthal angular distribution therefore provides a clear 
signal for the two gluon nature of the exchanged pomeron. 
This is supported by the fact that a very similar azimuthal 
distribution is obtained in the soft pomeron model by M.\ Diehl
\cite{Diehl,Diehl2}.
Figure \ref{fig:phidep} shows the $\phi$--dependence of the 
$ep$--cross section for the hard pomeron model, the soft pomeron 
model and for a boson gluon fusion process. 
We have normalized the cross section to unit integral to 
concentrate on the angular dependence. 
Thus a measurement of the azimuthal asymmetry of quark--antiquark 
jets will clearly improve our understanding of diffractive 
processes. 

Finally, we would like to mention the interesting issue of 
diffractive open charm production. 
It is in principle straightforward to extend our calculation to 
nonvanishing quark masses \cite{Lotter}.
A similar computation was done in \cite{NikZakcharm}. 
A more ambitious calculation has 
been performed in \cite{Durham} where also higher order 
correction have been estimated. 
The cross section for open charm production is again 
proportional to the square of the gluon density, the relevant 
scale of which is now modified by the charm quark mass
\be
d \sigma \sim \left[ \, x_{\Pam} G_p 
\left( x_{\Pam} ,(m_c^2 +\kf^2)\, \frac{Q^2+M^2}{M^2} 
\right) \right]^2 \,.
\ee
Integrating the phase space with the same cuts as above the 
charm contribution to the jet cross section is found to be \cite{Lotter}
\bea
  \sigma_{\mbox{\scriptsize tot}} &=& 8\,\mbox{pb} \phantom{0}\;\;\;\;\;
   \mbox{for} \; \kf^2 \ge 5 \,\mbox{GeV}^2 \nonumber\\
  \sigma_{\mbox{\scriptsize tot}} &=& 29\,\mbox{pb} \;\;\;\;\; 
   \mbox{for} \; \kf^2 \ge 2 \,\mbox{GeV}^2  \,. \nonumber
\eea
In the case of open charm production one can even integrate down to
$\kf^2=0$ since the charm quark mass sets the hard scale. One then 
finds for the $c\bar{c}$ contribution to the total diffractive cross section
\bea
  \sigma_{\mbox{\scriptsize tot}} &=& 101\,\mbox{pb} \phantom{0}\;\;\;\;\;
   \mbox{for} \; \kf^2 \ge 0 \,\mbox{GeV}^2 \;.\nonumber
\eea
In the soft pomeron model the corresponding numbers read \cite{Diehl}
\bea
  \sigma_{\mbox{\scriptsize tot}} &=& 4.8\,\mbox{pb} \;\;\;\;\;
   \mbox{for} \; \kf^2 \ge 5 \,\mbox{GeV}^2 \nonumber\\
  \sigma_{\mbox{\scriptsize tot}} &=& 17\,\mbox{pb} \phantom{.}\;\;\;\;\; 
   \mbox{for} \; \kf^2 \ge 2 \,\mbox{GeV}^2   \nonumber\\
  \sigma_{\mbox{\scriptsize tot}} &=& 59\,\mbox{pb} \phantom{.}\;\;\;\;\; 
   \mbox{for} \; \kf^2 \ge 0 \,\mbox{GeV}^2  \,. \nonumber
\eea
In \cite{Durham}, where corrections from gluon radiation 
are taken into account, 
the relative charm contribution to the
diffractive structure function is estimated to be of the order of
25--30\%.

\section{Summary and Outlook}
We have presented perturbative QCD calculations for the production 
of quark--antiquark jets in DIS diffractive dissociation. 
The results are parameter free predictions of the corresponding 
cross sections, available for light quark jets 
as well as for open charm production. 
The cross sections are proportional to the square of the gluon 
density and the relevant momentum scale of the gluon density 
has been determined. 
The azimuthal angular distribution of the 
light quark jets can serve as a clean 
signal for the two--gluon nature of the pomeron in hard processes. 

The two--jet final state is only the simplest case of 
jet production in diffractive deep inelastic scattering. 
The next steps should be the inclusion of order--$\alpha_s$ corrections
and the extension to processes with additional 
gluon jets. Such processes become dominant in the large--$M^2$ region.

\end{document}